\documentclass[a4paper,fleqn,usenatbib]{mnras}
\usepackage{newtxtext,newtxmath}
\usepackage[T1]{fontenc}
\usepackage{ae,aecompl}
\usepackage{graphicx,xcolor}
\usepackage{amsmath}
\usepackage{amssymb}
\title[Mechanism metre-wave stellar emission]{On the mechanism of polarised metrewave stellar emission}
\author[H. K. Vedantham]{H.~K.~Vedantham$^{1,2}$\thanks{E-mail: vedantham@astron.nl }
\\
$^{1}$ASTRON, The Netherlands Institute for Radio Astronomy, Oude Hogeveensedijk 4, 7991PD, Dwingeloo, The Netherlands\\
$^{2}$ Kapteyn Astronomical Institute, University of Groningen, Landleven 12, 9747 AD Groningen, The Netherlands
}
\date{Accepted XXX. Received YYY; in original form ZZZ}
\pubyear{2020}
\begin{document}
\label{firstpage}
\pagerange{\pageref{firstpage}--\pageref{lastpage}}
\maketitle
\begin{abstract}
Two coherent radio emission mechanisms operate in stellar coronae: plasma emission and cyclotron emission. They directly probe the electron density and magnetic field strength respectively. Most stellar radio detections have been made at cm-wavelengths where it is often not possible to uniquely identify the emission mechanism, hindering the utility of radio observations in probing coronal conditions. In anticipation of stellar observations from a suite of sensitive low-frequency ($\nu\sim 10^2\,{\rm MHz}$) radio telescopes, here I apply the general theory of coherent emission in non-relativistic plasma to the low-frequency case. I consider the recently reported low-frequency emission from dMe flare stars AD\,Leo and UV\,Ceti and the quiescent star GJ\,1151 as test cases. My main conclusion is that unlike the cm-wave regime, for reasonable turbulence saturation regimes, the emission mechanism in metre-wave observations ($\nu\sim 10^2\,{\rm MHz}$) can often be identified based on the observed brightness temperature, emission duration and polarisation fraction. I arrive at the following heuristic: M-dwarf emission that is $\gtrsim $\,hour-long with $\gtrsim 50\%$ circular polarised fraction at brightness temperatures of $\gtrsim 10^{12}\,$K at $\sim 100\,{\rm MHz}$ in M-dwarfs strongly favours a cyclotron maser interpretation. 
\end{abstract}
\section{Introduction}
\label{sec:introduction}
Several low-frequency ($\nu\sim 10^2\,{\rm MHz}$) observational efforts aim to study stellar coronal activity and exoplanetary magnetospheres \citep{lynch2017B,villadsen-2019,ogorman2018,turner2018,bastian2000,murphy2015,lynch2017A,vedantham-2020}. 
At metre-wavelengths, existing telescopes are only sensitive to radiation with very high brightness temperatures ($\gg 10^{10}\,{\rm K}$) that favour a coherent emission process. Observations of coherent emission are particularly constraining since they provides a clean measurement of the plasma density (in case of plasma emission) and magnetic field strength (in case of cyclotron emission) in the emitter.

Highly polarised radio bursts have been studied at cm-wavelength. The implied plasma densities and magnetic field strengths for the case of plasma and cyclotron emission are $\sim 10^{11}\,{\rm cm}^{-3}$ and $\sim 10^3\,$G, respectively. The emission site is likely to reside in coronal magnetic loops with heights significantly smaller than the stellar radius. A secure determination of the field strength and plasma density is however only possible if the emission mechanism can be uniquely identified. This is in general not straightforward; both emission mechanisms have been suggested to account for observations with comparable characteristics. \citep{zaitsev1983,stepanov-2001,melrose1982,villadsen-2019,slee-2003,zic-2019}

Metre-wavelengths are sensitive to much lower densities ($\sim 10^8-10^9\,{\rm cm}^{-3}$) and field strengths ($B\sim 10^2\,{\rm G}$) that are likely to persist at coronal heights comparable to or larger than the stellar radius. They, therefore, probe the tenuous higher coronal layers where one anticipates a transition between a radial coronal structure dictated by hydrostatic balance to one determined by the outflowing stellar wind, and occasionally, coronal mass ejections. Such data are of acute interest because they can be used to infer the space weather conditions around exoplanets.

Large metre-wavelength radio telescopes capable of mJy-level sensitivities are now available \citep{lofar,mwa,gmrt}. Coherent emission can attain brightness temperature of $\gtrsim 10^{12}\,{\rm K}$ which yields a distance-horizon of $\gtrsim 10\,$pc for mJy-level sensitivity--- sufficient for an unbiased survey of stars of varying spectral type and activity levels. In anticipation of a sumptuous yield of metre-wave stellar bursts (and perhaps exoplanetary emissions) from these surveys, I review the basic theory of coherent emissions and propagation, with a particular emphasis on metre-wavelengths (Section \S \ref{sec:mechanisms}). 

Targeted observations of anomalously active flare stars have recently detected several coherent radio bursts at metre-wavelengths. While such stars are not representative of the vast majority of quiescent stars, the observations provide a fortuitous opportunity to apply the basic principles of coherent emission in coronae. I undertake this in Section \S \ref{sec:observations}. 

The main conclusion I reach is that for $\nu\sim 10^2\,{\rm MHz}$, a brightness temperature that exceeds $\sim 10^{12}\,{\rm K}$ and high polarisation fraction ($\gtrsim 50\%$) favours a cyclotron maser interpretation for prototypical of M-dwarf coronal parameters. In some highly active stars, the increases coronal temperature and density scale height might lead to highly polarised plasma emission reaching a brightness temperature of $\sim 10^{12}\,{\rm K}$.
On the other hand, in the cm-wave regime, the intensity of observed long-duration emission can be explained by both cyclotron maser and plasma emission. Hence low-frequency observations of stellar bursts will usually lead to a clear identification of the emission mechanism.

A glossary of symbols and their meaning is given in the Appendix for quick reference. I use Gaussian units unless specified otherwise. 

\section{Coherent emission Mechanisms}
\label{sec:mechanisms}
There are two known coherent emission mechanisms in non-relativistic plasma: cyclotron emission and plasma emission. While reviewing the necessary theory here, I only present the basic equations and first-principles here, and refer the reader to the original sources \citet{benz-book,melrose1982,zaitsev1983,wu1979,treumann} for further details.

\subsection{Cyclotron maser}
\label{subsec:ecmi_emission}
Cyclotron maser emission occurs at the fundamental and harmonics of the cyclotron frequency: $\nu_c\approx 2.8\times 10^6\,B$, where $B$ is the ambient magnetic field strength. The source of free energy in cyclotron maser is an inverted (or unstable) population of electrons. In the two-dimensional momentum space (axis parallel and perpendicular to the magnetic field), the inversion usually takes the form of either a loss-cone or horseshoe-shaped distribution\footnote{Loss cone refers to an absence of gyrating electrons at low pitch angles creating a population inversion in perpendicular (to the B-field) momentum space, that drives the maser.} of {\em mildly relativistic} electrons ($E\gtrsim 10\,{\rm keV}$). 
The horseshoe-type maser can generate brightness temperatures that are orders of magnitude larger than that from a loss-cone distribution, but in general,   requires an accelerating electric field that is parallel to the magnetic field. It is presently unclear if such an electric field can be sustained over long periods in a dense corona which can readily shield such fields. 
The loss-cone, on the other hand, can be readily established via magnetic mirroring in a magnetic trap such as a coronal loop or the large-scale magnetic field of an object.  
Consequently, horseshoe masers are generally thought to drive auroral emissions in the rarefied magnetospheres of the planets, and possibly some short-duration ($\ll 1\,{\rm sec}$) `spike-bursts' in stellar coronae. Because, we are only considering long-duration emission here, we assume that any cyclotron maser emission from corona must be of the loss-cone type.

The theory of loss-cone masers has been studied extensively. Here we adopt expressions from \citet{melrose1982} for the radiation brightness temperature.  Let the maser be driven by a non-thermal population of mildly-relativistic electrons with a density $n_o$ and characteristic speed $v_0$. The maximum attainable brightness temperature in a loss-cone maser is 
\begin{equation}
    T_{\rm b}^{\rm max} \approx \frac{1}{2\pi}\left(\frac{n_om_ev_o^2}{k_{\rm B}}\right)\left(\frac{c^2}{\nu v_o}\right)^3.
    \label{eqn:cycl_tbmax}
\end{equation}
The first term within parentheses is the kinetic energy density in the emitting electrons which for $v_o=0.2c$ is $\sim n_o 10^8\,{\rm K}$. The peak brightness temperature is therefore
\begin{equation}
    T_{\rm b}^{\rm max} \approx 5\times 10^{16} n_0\beta_{0.2}^{-1}\nu_{100}^{-3}.
    \label{eqn:cycl_tbmax1}
\end{equation}
The finite rate at which the loss cone forms and empties reduces the brightness temperature of the time-averaged emission by a factor of $v_o/(\Gamma L)$ where $\Gamma$ is the growth-rate of the maser, $L$ is the length-scale of the magnetic trap ($v_o/L$ is the rate at which the loss-cone forms). The resulting time-averaged brightness temperature is 
\begin{equation}
    T^{\rm avg}_{\rm b} \approx \frac{m_ev_0^2}{4\pi k_B}\frac{c^2}{\nu^2Lr_0},
    \label{eqn:cycl_tbavg}
\end{equation}
where $r_0$ is the classical electron radius. In convenient units, we have 
\begin{equation}
    T_{\rmn B}^{\rm avg}\approx 10^{14.8}\,\beta_{0.2}^2         \nu_{100}^{-2}\left(\frac{L}{10^{10}\,{\rm cm}}\right)^{-1}.
\label{eqn:cycl_tbavg1}
\end{equation}
Individual spikes can attain brightness approaching $T_{\rm b}^{\rm max}$, but only with a small duty factor. The time-averaged brightness temperature cannot exceed $T_{\rm b}^{\rm avg}$.

The polarisation of cyclotron maser emission is $\approx 100\%$ in the $x$- or $o$-mode depending on whether emission at the fundamental or the second harmonic has the higher growth-rate allowing it to extract most of the available free energy.\footnote{The two nature modes in a magnetised plasma are elliptically polarised with axial ratios of $T_x$ and $T_o=-1/T_x$. $x$-mode has an electric field rotating in the same sense as the gyrating electrons.} In a loss-cone driven maser, for $\nu_{\rm c}/\nu_{\rm p}\lesssim 0.2$, $x$-mode at the fundamental cyclotron frequency is expected to dominate, whereas $o$-mode at the fundamental or the $x$-mode at the second harmonic will dominate at higher ratios.

Cyclotron maser adequately describes the emission properties of all magnetised solar system planets. Although such emission must occur in stellar coronae, cyclotron absorption by dense ambient thermal plasma in stellar coronae require one to postulate an additional mechanism to ensure radiation escape. Because cyclotron absorption falls rapidly for small angles between the magnetic field and the wave vector (details in Appendix A), the proposed mechanisms invariable involve guiding the emitted energy along magnetic field lines, via some scattering mechanism \citep[see for e.g.][]{zaitsev2005} or via wave refraction \citep[see for e.g.][]{speirs-2014}.

To convert observed flux densities to brightness temperature, we need the transverse emitter size. This must be determined jointly by (a) the beaming properties and instantaneous bandwidth of the maser and (b) the `active' section of the source to which the mildly relativistic electrons driving the maser are confined. The latter especially depends on the electron acceleration mechanism which is not known apriori. For simplicity, we assume that the emitter is a circle of radius $xR_\ast$ where $R_\ast$ is the physical radius of the emitting object. If $F$ is the observed flux density, then the observationally inferred brightness temperature is 

\begin{equation}
    T_{\rm b}^{\rm obs} = \frac{F\lambda^2}{2k_{\rm B}}  \frac{d^2}{\pi x^2R_\ast^2} \approx 10^{13}\frac{F}{\rm mJy} \nu_{100}^{-2}x^{-2}R_{\ast,10}^{-2}d_{10}^2\,\,{\rm K}.
    \label{eqn:cycl_tbobs}
\end{equation}

Comparing this with the theoretical estimate, we find that loss-cone driven masers on stars at $d\sim 10{\rm pc}$ can reach $\sim {\rm mJy}$ level flux-densities at 100\,MHz.
\subsection{Plasma emission}
\label{subsec:plasma_emission}
As before, I present the final equations and basic arguments here and refer the reader to Appendix B and the original sources \citep{benz-book,stepanov-2001,zaitsev1983,stepanov-1999} for further details.
Plasma emission occurs at the fundamental or harmonic of the plasma frequency: $\nu_p\approx 10^4n_e^{0.5}$. The energy source for coronal plasma emission is the turbulent injection of impulsively heated plasma into an ambient cooler and denser medium. Such a situation can likely persist in confined plasma in flaring coronal loops. Emission at the fundamental plasma frequency is generated by scattering of plasma density waves (called Langmuir waves) on thermal ions. Emission at the second harmonic is generated by coalescence of two Langmuir waves travelling in opposite directions. 

Detailed expressions for the relevant emission and absorption coefficient have been derived elsewhere. Here I use the expressions for emergent brightness temperature given by \citet{stepanov-2001} that include the effects of spontaneous emission along with the non-linear effects of absorption by collisional damping and induced emission. At very low frequencies and for typical density scale heights in M-dwarfs coronae, the brightness temperatures derived by assuming spontaneous emission in an optically thin source are good approximations (as will be evident later). I, therefore, present these approximate equations here (further details in the Appendix B) while cautioning the reader that they may not be applicable at cm-wavelengths. 

The brightness temperature for optical thin fundamental and second harmonic plasma emission are given by 
\begin{equation}
\label{eqn:plasma_tbfmax}
    T_{\rm b, f} \sim 2\times 10^{10} \nu_{100} \left(\frac{h_{\rm p}}{10^{10}\,{\rm cm}}\right) \left(\frac{T_1}{10^8\,{\rm K}}\right) \left(\frac{w}{10^{-5}}\right)
\end{equation}
and
\begin{equation}
\label{eqn:plasma_tbhmax}
    T_{\rm b, h} \sim 4\times 10^{10} \left(\frac{h_{\rm p}}{10^{10}\,{\rm cm}}\right) \left(\frac{T}{10^6\,{\rm K}}\right)^4
    \left(\frac{T_1}{10^8\,{\rm K}}\right)^{-1/2}
    \left(\frac{w}{10^{-5}}\right)^2
\end{equation}
The brightness temperatures cannot be arbitrarily large and will saturate at some level. At metre-wavelengths, harmonic emission dominates and is most likely to saturate \citep{stepanov-2001}. This is especially true in flare stars with hot coronae because $T_{\rm b,h}\propto h_{\rm p}T^4\propto T^5$. Saturation happens when the brightness temperature of emission equals that of the Langmuir waves \citep{melrose1980} given by 
\begin{equation}
\label{eqn:tb_max_lang}
    T^{\rm max}_{\rm b,h}\sim 10^{12}\left(\frac{\nu_{\rm p}}{100\,{\rm MHz}}\right)^{-1} \left( \frac{T}{10^6\,{\rm K}}\right)^{3/2}\left(\frac{T_1}{5\times 10^7\,{\rm K}}\right)
\end{equation}

The beaming pattern of plasma emission at both the fundamental and harmonic is broad. Because a stellar disk may not be filled entirely by flaring loops, I take the transverse area of the emitter to be a fraction $f$ of the stellar disk. 
The observed flux density then has a simple relationship to the brightness temperature of the emitter:
\begin{eqnarray}
    T_{\rm b}^{\rm obs} &=& \frac{F\lambda^2d^2}{2k_{\rm B}f\pi R_\ast^2}\nonumber \\ &\approx& 10^{14}\nu_{100}^{-2}d_{10}^2R_{\ast,10}^{-2} \left(\frac{f}{0.1}\right)^{-1}\left(\frac{F}{\rm mJy}\right)\,\,{\rm K}
    \label{eqn:tobs_plasma}
\end{eqnarray}

Even if $f\approx 1$ (entire disk is filled with flaring loops), it is clear from equations \ref{eqn:tobs_plasma} and \ref{eqn:plasma_tbfmax} that fundamental plasma emission is unlikely to generate a $\sim {\rm mJy}$ level source at $d\sim 10\,{\rm pc}$ at $\nu\sim 100\,{\rm MHz}$. However in much hotter coronae with $T\approx 5\times 10^6\,{\rm K}$, a comparison of equations \ref{eqn:tobs_plasma} and \ref{eqn:plasma_tbhmax} shows that the second harmonic emission may generate a $\sim {\rm mJy}$ level source with the above assumptions. 

\subsubsection{Polarisation of plasma emission}
\label{subsec:plasma_pol}
Fundamental plasma emission is $\approx 100\%$ polarised in the $o$-mode.
Second harmonic plasma emission from solar bursts are only seen to have moderate polarisation fractions of $\lesssim 20\%$, but this is likely due to lower magnetic field strength in the solar corona. At higher magnetic field strengths which are likely to be encountered in coronae of late-type stars, the theory allows higher polarisation fractions.

Second harmonic emission can be polarised in either $x$- or $o$-mode depending on whether the Langmuir wave spectrum is isotropic or preferentially beamed. The polarisation fraction in the two cases is $\approx 1.8|\cos\theta|\nu_{\rm c}/\nu_{\rm p}$ and $0.23|\cos\theta|\nu_{\rm c}/\nu_{\rm p}$ respectively, where $\theta$ is the angle between the wave vector and the magnetic field. These are approximations valid for $\nu_{\rm c}\ll\nu_{\rm p}$, which is likely true in the solar corona but not necessarily the case in other stars, especially M-dwarfs with strong magnetic fields. \citet{melrose-2harmpol} have derived analytic expressions for the general case, which I have computed and plotted in figure \ref{fig:2harmpol} for the case of isotropic Langmuir wave spectrum (see Appendix C for details). The figure shows that angle averaged polarisation fractions of up to $\approx 75\%$ in the $x$-mode are feasible for $\nu_{\rm p}/\nu_{\rm c}\approx 0.6$. At lower ratios, the polarised fraction drops quasi-linearly and at higher ratios, gyro-harmonic absorption at the $s=3$ layer allows radiation escape only in contrived geometries (see Appendix A for details). 

\begin{figure}
    \centering
    \includegraphics[width=\linewidth]{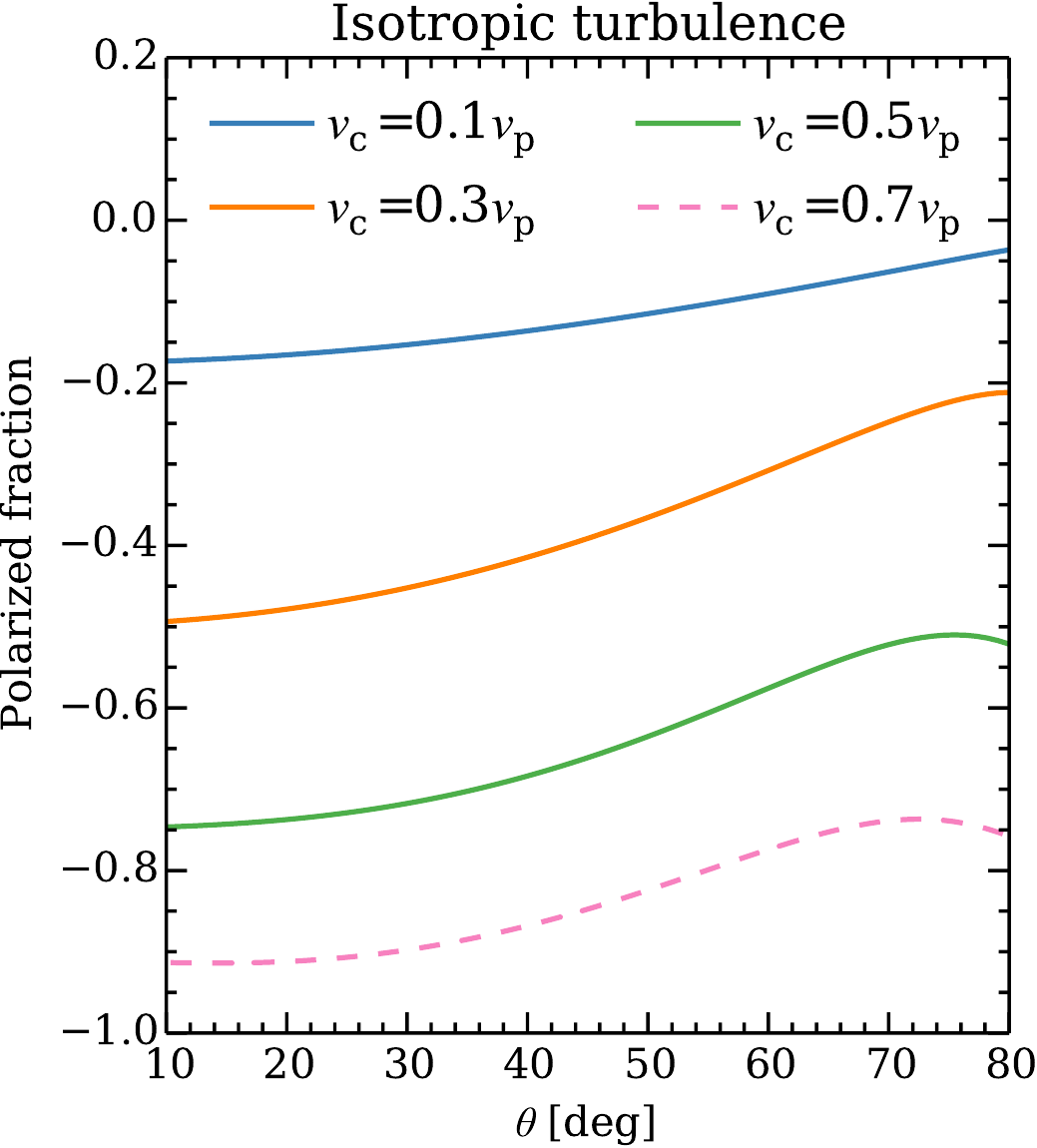}
    \caption{Polarization of second harmonic emission excited by an isotropic distribution of Langmuir waves as a function of the angle between wave-vector and the magnetic field. The curves were computed using the analytic expression given by \citet{melrose-2harmpol}. The curves are for different values of plasma to cyclotron frequency ratio. Negative polarization indicates $x$-mode dominated emission. The dotted line implies that radiation escape is only possible in contrived geometries (for $\nu_{\rm c}/\nu_{\rm p}\gtrsim 0.6$; see Appendix A\&C for details).}
    \label{fig:2harmpol}
\end{figure}

The observed polarisation of second harmonic radiation over a broad bandwidth will be lower because the ratio $\nu_{\rm p}/\nu_{\rm c}$ will invariably vary within coronal loops. 
Additionally, the observed polarised fraction of the fundamental and the harmonic will be lowered due to superposition of emission of opposite polarity from regions of positive and negative line of sight magnetic field components. 
If the size of the emitter is a large fraction of the surface area of the star, then in the absence of a special viewing geometry, the emission will be a nearby equal mix of waves with opposing polarities and the aggregate radiation will be unpolarised. 
As such, low polarisation fractions are expected if the size of the emitter is a substantial fraction of the stellar disc.

Further care is warranted in interpreting observational biases in polarisation fraction, particularly at metre-wavelengths where growth-rates for the harmonic are very high. Fundamental plasma emission and cyclotron emission are 100\% polarised, although the observed fraction may be smaller due to geometric and propagation effects. Second harmonic emission, on the other hand, requires somewhat contrived circumstances to achieve high polarisation fractions: the cyclotron frequency should be high enough but not too high to prevent the escape of radiation. Hence if a survey selects for a high polarisation fraction, it is conceivable that second harmonic bursts with high polarisation fractions are detected, albeit in a narrow frequency range. However, in a polarisation-fraction blind survey, a prototypical second harmonic burst is unlikely to be highly polarised.  

\subsection{Bandwidth and duration}
\label{subsec:bandwidth_duration}
Both plasma and cyclotron emission are inherently narrowband ($\Delta\nu/\nu \lesssim 0.01$) but coronae and magnetosphere are known to have a wide range of density and magnetic field variation. Hence emission bandwidth itself is not a clear discriminant between emission mechanisms. 

Cyclotron maser has extremely high growth rates ($\sim 10^{-4}\nu_{\rm c}\,{\rm s}^{-1}$ which can lead to wave growth from the thermal level of the plasma ($\sim 10^8\,{\rm K}$ typically) to brightness temperatures of $10^{15}\,{\rm K}$ (i.e. about 17 $e$-foldings) in just $\sim 2\,{\rm ms}$. Observation of temporal structure on ms-timescales almost always implies a cyclotron maser origin. 

The duration of cyclotron maser emission is largely a geometric effect. The emission from a given site is beamed along the surface of a cone whose axis is parallel to the ambient magnetic field, which has a large opening angle and small thickness, although propagation effects can alter this simple geometry. In addition, depending on the mechanism of electron acceleration, the emitting electron may be restricted to a small azimuthal sector. This is the case when the emission is powered via electrodynamic interaction with an orbiting satellite. On the other hand, if the emission is powered by the breakdown of co-rotation between the plasma and the co-rotating magnetic field, then the emitting electron occupy all azimuthal angles. As such, depending on the rotation period of the star and the orbital period of planets that may be inducing the emission, the duration of over which the emission cone points to an observer can vary widely from minutes to several hours or even days.

Langmuir waves leading to plasma emission can grow over millisecond timescales \citep[see for e.g. eqn. 7 of][]{zaitsev1983}, but the conversion of Langmuir waves to transverse electromagnetic waves sets the bottleneck in extracting short-duration bursts via the plasma mechanism. Consider an instantaneous injected of supra-thermal electrons into the corona. Initially, the Langmuir waves are directed along a narrow forward-cone and only fundamental plasma emission is feasible. The growth rates of emission at metre-wavelengths (see equations \ref{eqn:alpha_f} and \ref{eqn:alpha_h}) is $\sim 10\,{\rm K/cm}$. Hence $10^{12}\,{\rm K}$ of brightness temperature can build over several seconds. This sets the minimum time-scale of variability to the order of several seconds for fundamental plasma emission. There is no strict upper limit to the duration of plasma emission. So long as turbulent energy is injected into the corona, the emission will persist.

The treatment of plasma emission presented here assumes a well formed Langmuir wave turbulence that is isotropic and has a flat spectrum. At the point of injection of hot plasma, the Langmuir waves are instead confined to a narrow forward cone. Angular scattering of Langmuir waves proceeds on a timescale, $\tau_{\rm L}$, of \citep[eqn 3.12 of ][where $\tau_L \sim W_k/J_k$ in their notation]{kaplan1973}
\begin{equation}
\label{eqn:taul}
\tau_{\rm L} \sim \nu_{100}^{-2}T_6^{1.5}\,{\rm min}
\end{equation}
I note the distinction between short-duration events and long-duration events based on the nature of turbulence as applied to our calculation of brightness temperature in \S\ref{subsec:plasma_emission}. I have assumed here that the Langmuir-wave spectrum is isotropic. If instead, the incident turbulent energy is concentrated into a narrow cone of opening angle $\Omega_L$, then the spectral energy density of Langmuir waves (and hence their temperature) increases by a factor of $4\pi \Omega_L^{-2}$ compared to the isotropic case for the same overall energy density. Hence on timescales comparable to of less than $\tau_{L}$ the bounds on brightness temperature computed in \S\ref{subsec:plasma_emission} can be readily exceeded by this factor. Therefore, at metre-wavelengths, I make a distinction between short bursts (on minute timescales or less) and long bursts (on tens of minutes timescales and longer). The brightness temperature limits of \S\ref{subsec:plasma_emission} only strictly apply to the latter.

\section{Discussion}
\label{sec:observations}
\subsection{Application to metre-wave detections}
I now apply the the theory developed in \S\ref{sec:mechanisms} to metre-wave observations. Details of the observations are given in Table \ref{tab:obs_details}. Long-duration polarised bursts from AD\,Leo were reported at $\sim 300\,{\rm MHz}$ by \citep{villadsen-2019}, whereas \citet{lynch2017A} reported bursts from UV\,Ceti at 150\,MHz. Both of these are highly active flare stars. As a contrast, I also consider the long-duration polarised emission from the quiescent star, GJ\,1151, reported by \citet{vedantham-2020}. As a reference value, the table lists the `observed' brightness temperature assuming that the source is a filled disc with a radius given by the stellar radius. I have used the empirical scaling laws of \citet{johnstone2015} to convert the observed X-ray luminosity of the stars (upper limit in case of GJ\,1151) to the temperature of the thermal plasma. 

\begin{figure}
\centering
\includegraphics[width=\linewidth]{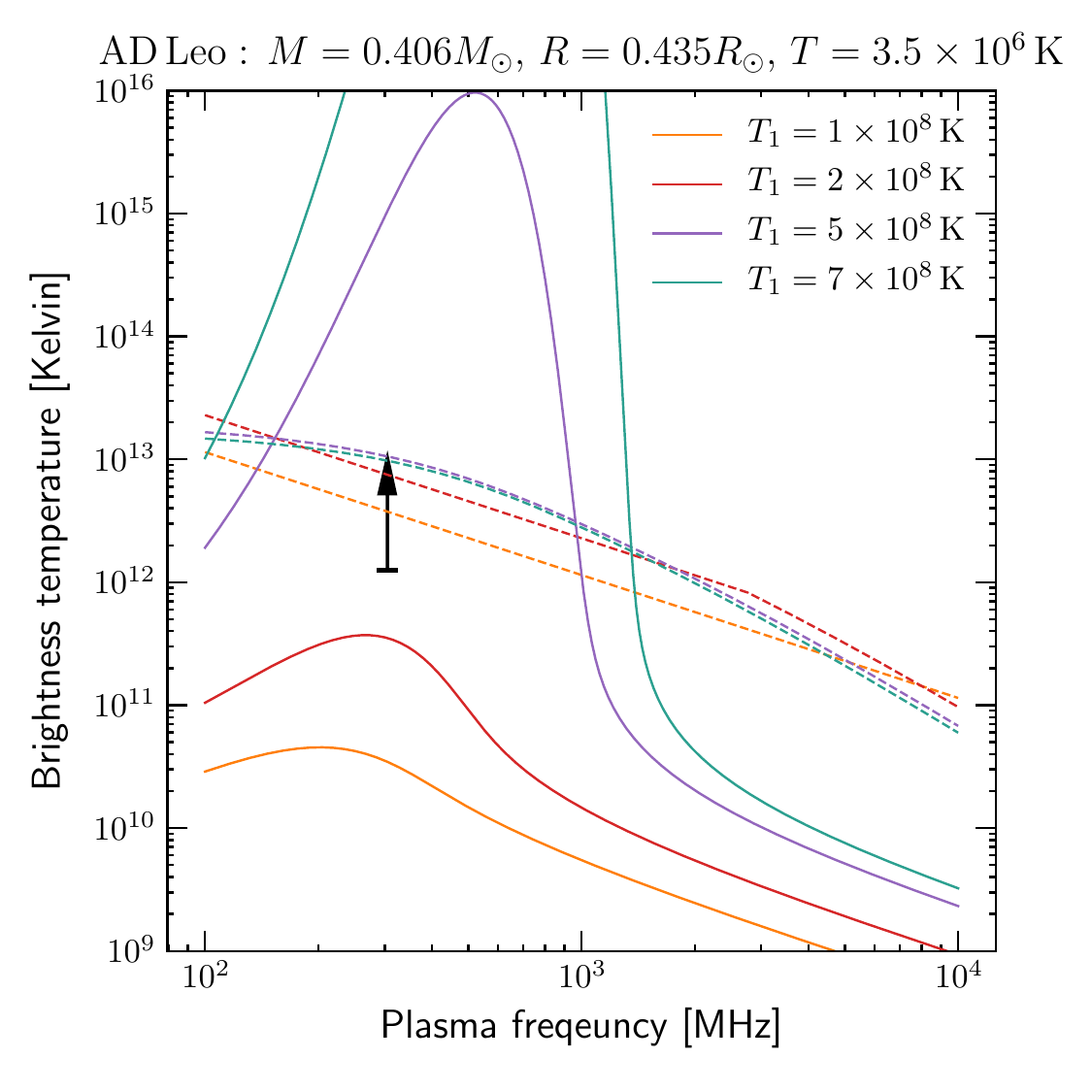}
\caption{Peak radiation brightness temperature of fundamental (solid lines) and second harmonic (dashed lines) plasma emission for parameters relevant to AD Leo. The different curves are for different temperatures of the hot plasma component. The black arrow square shows the observationally inferred lower limit computed by assuming that the transverse size of the emitter is the same as that of the stellar disk. The black bar at the base of the arrow spans the observed bandwidth of emission. The quasi-linear rise  of the fundamental curve low-frequencies is consistent with equation \ref{eqn:plasma_tbfmax}. The approximately flat profile of the second-harmonic curve at low frequencies is consistent with equation \ref{eqn:plasma_tbhmax}. \label{fig:tb_plasma_adleo}}
\end{figure}

\begin{figure}
\centering
\includegraphics[width=\linewidth]{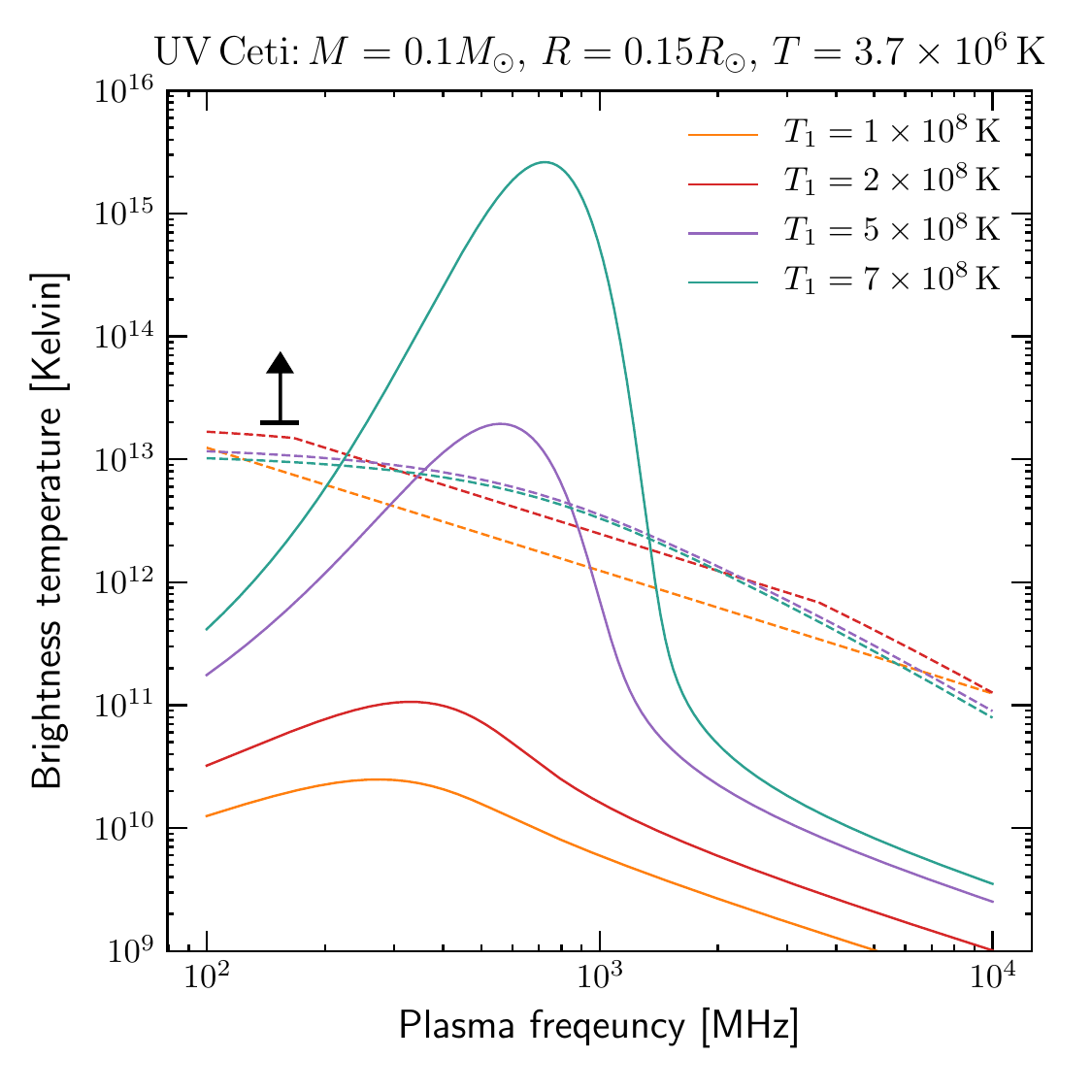}
\caption{Same as figure \ref{fig:tb_plasma_adleo} but for parameters relevant to UV Ceti. Fundamental plasma emission can only account for the observed radio emission if the entire stellar disk has active coronal loops with a saturated level of turbulence. \label{fig:tb_plasma_uvceti}}
\end{figure}

\begin{figure}
\centering
\includegraphics[width=\linewidth]{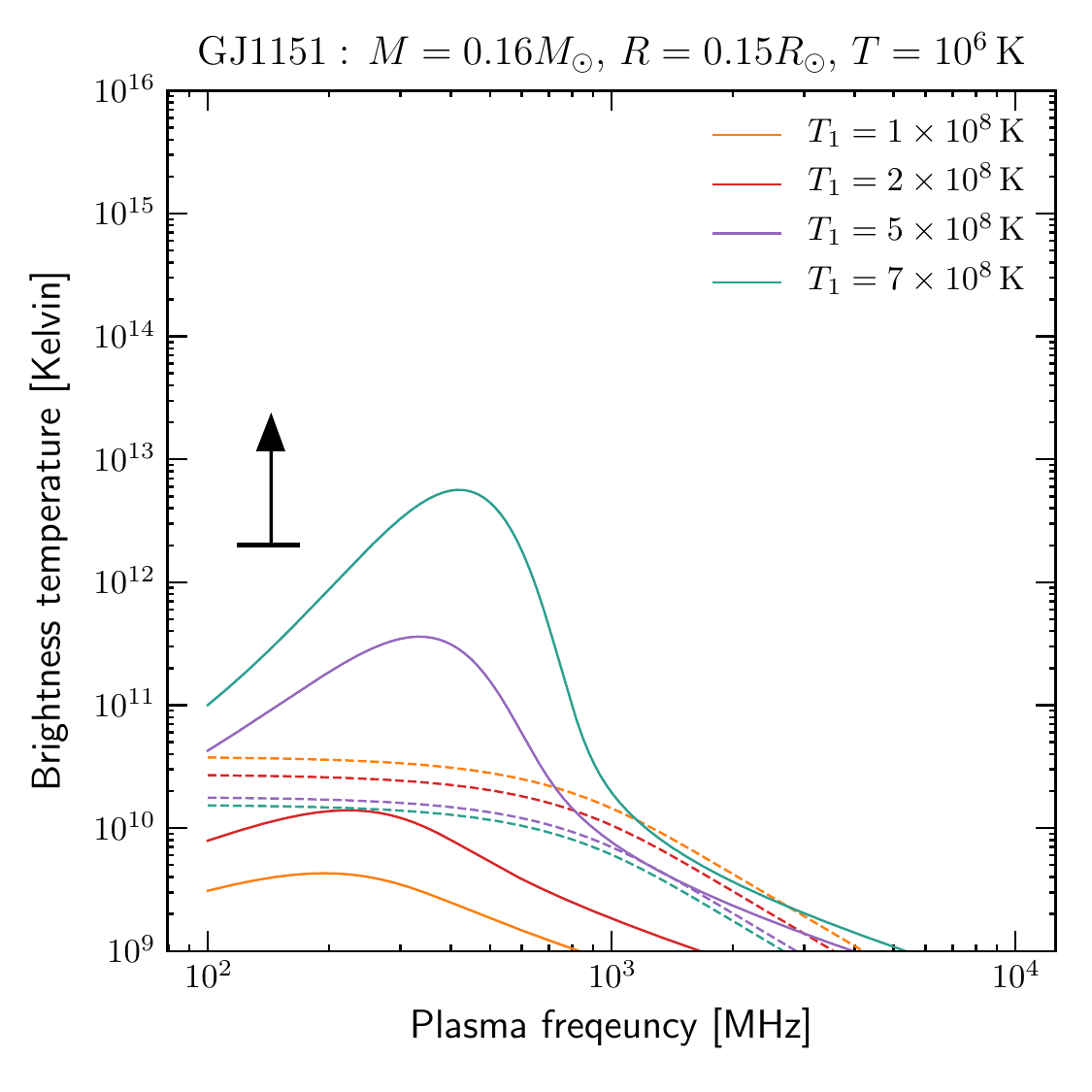}
\caption{Same as figure \ref{fig:tb_plasma_GJ1151} but for parameters relevant to GJ\,1151. The low coronal density scale height, as implied by X-ray luminosity limits, makes the fundamental plasma emission hypothesis untenable. \label{fig:tb_plasma_GJ1151}}
\end{figure}

In figures \ref{fig:tb_plasma_adleo}, \ref{fig:tb_plasma_uvceti} and \ref{fig:tb_plasma_GJ1151}, I have plotted the maximum brightness temperature feasible from fundamental and second harmonic plasma emission for the three stars as a function of the emission frequency and the temperature of the impulsively heated hot plasma component. The curves are computed using equations presented by \citet{stepanov-2001} and share a characteristic profile. The declining efficient of harmonic emission at higher frequencies and the increased collisional damping rate at higher frequencies lead to a monotonically decreasing curve for the harmonic component. The fundamental component rises quasi-linearly at low frequencies due to spontaneous emission as approximated by the relationships in \S \ref{sec:mechanisms}. At frequencies around 1 GHz, it grows non-linearly due to induced emission, and eventually declines at higher frequencies as the collisional damping rate overcomes the rate of induced emission. I note here that the equations of \citet{stepanov-2001} for the fundamental component only consider the conversion of Langmuir to electromagnetic waves. They do not include the inverse process. In particular, the induced conversion of electromagnetic waves to Langmuir waves will ultimately limit the temperature of the electromagnetic waves. A detailed discussion of the highest brightness temperature possible for fundamental plasma emission is unnecessary for our purposes here. Suffice it to say that the brightness temperatures we are dealing with here are well within the limiting value \citep[see e.g. ][]{stepanov-2001}.

\begin{table*}
\label{tab:obs_details}
    \centering
    \begin{tabular}{l|l|l|l|l}
         {\bf Parameter} &  {\bf AD\,Leo} & {\bf UV\,Ceti}  & {\bf GJ\,1151} & {\bf Data source}\\
         Sp. type & M3.4 & M6  & M4.5 & (S1,S4,S1)\\
         Distance/pc &4.97 & 2.69  & 8.04 & (S2,S2,S2)\\
         Mass$/M_\odot$ & 0.406 & 0.1  & 0.154 & (S1,S5,S1)\\
         Radius$/R_\odot$ & 0.435 & 0.15  & 0.19 & (S1,S5,S1)\\
         $L_X$\,[ergs/s] & $10^{28.92}$ & $10^{27.1}$  & $<10^{26.2}$ & (S3,S6,S8)\\
         $F_X$\,[ergs/s/cm$^{2}$] & $10^{5.75}$ & $10^{5.86}$  & $<10^{4.88}$ & $=L_X/(4\pi R_\ast^2)$\\
         $T_{\rm cor}\,[10^6\,{\rm K}]$ & 3.5 & 3.7 & $<2$ & $=0.11F_X^{0.26}$\\
         $h_p$ & $0.324R_\ast$ & $0.48R_\ast$& $<0.2R_\ast$ & \\
         ${\rm log}_{10}T_{\rm b}$ (observed$^\dagger$) & 12.8, 12.1 & 12.5, 13.3  &  12.3 & (S3,S7,S9)\\
         Pol. frac & $>96\%,\,>87\%$&$>27\%$ & $64\pm6\%$ & (S3,S7,S9)\\
         Polarity (LCP/RCP) & Unknown & Both &  RCP & (S3,S7,S9)\\
         Duration & $>3.5^{\rm h}$ & $0.5^{\rm h}$ & $>8^{\rm h}$ & (S3,S7,S9)\\
         ${\rm log}_{10}L_R\,$[ergs/s/Hz] & 14.28& 14.91 & 13.8 & (S3,S7,S9)\\
         Visible pole & S & N & Unknown & (S3,S7,S9)\\
    \end{tabular}
    \caption{Relevant properties of the corona and radio emission of the stars considered here.  The data sources are as follows. S1: \citet{mann2015}, S2: \citet{gaia-dr2}, S3: \citet{villadsen-2019}, S4: \citet{kirk1991}, S5: \citet{uvceti-mass}, S6: \citet{audard2003}, S7: \citet{lynch2017A}, S8: \citet{wright-2018}, S9: \citet{vedantham-2020} $^\dagger$Observed $T_{\rm b}$ is computed with a projected source size $=\pi R_\ast^2$.}
\end{table*}

\subsubsection{AD\,Leo}
\label{subsec:obs_adleo}
\citet{villadsen-2019} reported two bursts from the flare star AD Leo between $290$ and $320\,{\rm MHz}$ which flux-densities of 38 and 210\,mJy. The brighter of the two bursts only lasted about 5 minutes and I classify this as a short-duration burst vis-a-vis plasma emission (see \S \ref{subsec:bandwidth_duration}). The second, weaker burst which is listed in table \ref{tab:obs_details} lasted more than $3.5^{\rm h}$ and is a long-duration event. Although weaker, the latter provides a cleaner constraint on plasma emission models because an isotropic spectrum of turbulence can be assumed while computing the spectral density of Langmuir waves.

For the long-duration event, the high polarization fraction rules out second harmonic emission, which leaves us with fundamental plasma emission and cyclotron emission. From figure \ref{fig:tb_plasma_adleo}, it is evident that a hot component temperature of $T_1\sim 10^{8.6}\,{\rm K}$, can account for the observed emission even if only a fraction of the stellar disk is filled with flaring coronal loops emitting fundamental plasma emission. The observed brightness temperature can be trivially supported by a loss-cone cyclotron maser (see eqn. \ref{eqn:cycl_tbavg1}). Therefore, for this star, the emission mechanism for the observed 300\,MHz emission cannot be readily identified based on the brightness temperature and polarised fraction.

\subsubsection{UV Ceti}
\label{subsec:obs_uvceti}
\citet{lynch2017A} detected bursts from UV Ceti only in polarized emission providing a lower limit on the polarized fraction of about $27\%$ which is insufficient to rule out second harmonic plasma emission. Hence both fundamental and harmonic plasma emission, as well as cyclotron-maser emission must all be considered.
Figure \ref{fig:tb_plasma_uvceti} shows that fundamental plasma emission is untenable. Second harmonic plasma emission is feasible only if the entire stellar disk is filled by hot plasma with temperature $T_1\gtrsim 7\times 10^8\,{\rm K}$ at the theoretical maximum level of turbulence.  This is unprecedented as it requires the entire surface of the star to flare simultaneously instead of a few active coronal loops as seen on the Sun. I therefore conclude that the emission from UV Ceti is driven by cyclotron maser.

\subsubsection{GJ 1151}
\citet{vedantham-2020} present long-duration polarised emission from the quiescent star GJ\,1151 at 150\,MHz. GJ\,1151 provides an immediate contrast because it is a quiescent star with a low (thus far undetected) X-ray emission. The observed polarisation fraction rules out second harmonic plasma emission. 
Figure \ref{fig:tb_plasma_GJ1151}  shows that even in a contrived case where the entire surface of the star is filled with simultaneously flaring coronal loops, plasma emission cannot supply the observed brightness temperature. The low frequency of observation and the high observed brightness temperature, therefore, lead to a cyclotron maser interpretation for GJ\,1151.

The large impact of GJ1151's quiescence on the peak plasma brightness temperature estimates is noteworthy. This is apparent if one compares the curves in Fig. \ref{fig:tb_plasma_GJ1151} and those in \ref{fig:tb_plasma_uvceti} and \ref{fig:tb_plasma_adleo}. The reason for this disparity is the low coronal temperature of GJ1151 (determined from its X-ray faintness). A low coronal temperature leads to lower emissivity at the harmonic (see Eqn. \ref{eqn:plasma_tbhmax}), and a lower density scale height over which the emission at either the fundamental of the harmonic can grow. The disparate scale height leads to large disparities in fundamental emission at higher frequencies ($\nu\sim 1\,{\rm GHz}$) where induced emission (maser effect) becomes dominant.

\subsubsection{EQ Peg}
\citet{villadsen-2019} report a 0.27\,Jy burst from EQ Peg at 350\,MHz which is $>70\%$ circularly polarised. EQ Peg has stellar parameters similar to the UV Ceti system, using which, we get $T_{\rm b}\approx 10^{14}\,{\rm K}$. The similarity with UV Ceti also implies that Fig. \ref{fig:tb_plasma_uvceti} is applicable to UV Ceti which clearly shows that plasma emission cannot account for the observed brightness temperature. The emission from EQ Peg observed by \citet{villadsen-2019} is therefore due to cyclotron maser emission. 

\subsection{Comparison to cm-wave detections}
\citet{villadsen-2019} also report higher frequency long-duration emission from AD Leo and UV Ceti. The brightest of these reach 64\,mJy between 1 and 1.6\,GHz on AD Leo, and 30\,mJy between 1 and 6 GHz on UV Ceti. The peak brightness temperatures of the bursts on the two stars are $T_{\rm b}\approx 10^{11.2}\,{\rm K}$. Figures \ref{fig:tb_plasma_adleo} and \ref{fig:tb_plasma_uvceti} shows that this brightness temperature can be readily achieved with fundamental plasma emission even if an isolated active region on the star is responsible for the radio emission. The intensity of emission and the polarisation fraction alone cannot be used to distinguish between the two plausible mechanisms.

\subsection{Conclusions \& Outlook}
\label{subsec:conclusion}
New stellar radio observations have identified long-duration metre-wave bursts of coherent emission. Coherent plasma emission and cyclotron maser directly probe the ambient electron density and magnetic field. As such the metre-wave data can provide an understanding of of the component of dMe coronae that is much more tenuous that the component probed by cm-wave observations and X-ray data.

I have endeavoured to identify the origin of metre-wave emission observed from observations of the flare stars AD Leo and UV Ceti and the quiescent star GJ\,1151. My over-arching conclusion is that unlike the cm-wave regime, at sufficiently low frequencies ($\nu\sim 10^2\,{\rm MHz}$), brightness temperatures exceeding $\sim 10^{12}\,{\rm K}$ over hour-long duration are unlikely to be generated by fundamental plasma emission for the vast majority of M-dwarf coronal parameters. This allows us to identify the emission mechanism--- the first and necessary step in determining coronal parameters.

While this paper has brought the polarised fraction, brightness temperature, and temporal duration of emission to bear, the sense of polarisation ($x$- or $o$-mode) has not been considered as an identifier of the emission mechanism. This will require a treatment of the beaming properties of plasma and cyclotron maser emission as well as wave propagation effects within coronae; I leave this for future work.

\section*{Acknowledgements}
I thank Prof. Gregg Hallinan and Prof. Robert Bingham for discussions,  Dr Joe Callingham for commenting on an earlier version of the manuscript, and the anonymous referee for catching a crucial mistake in my brightness temperature calculations.

\appendix

\begin{table*}
    \centering
    \caption{Glossary of symbols and their meaning.}
    \begin{tabular}{l|l}
         $R_\ast$&Stellar radius   \\
         $R_{\ast,10}$&$R_\ast/(10^{10}\,{\rm cm})$   \\
         $d$ &Distance to the star \\
         $d_{10}$&$d/(10\,{\rm pc})$ \\
         $ B$ & Magnetic field strength \\
         $n_0$& Hot plasma density \\
         $\nu$&Spectral frequency \\
         $\nu_{100}$&$\nu/(100\,{\rm MHz})$ \\
         $\omega$&Angular frequency $=2\pi\nu$\\
         $\lambda$ & Wavelength \\
         $k$ & Wave-vector $=2\pi/\lambda$\\
         $\nu$ & Electromagnetic wave frequency\\
         $\Delta \nu$ & Emission bandwidth\\
         $\nu_{\rm p}$&Plasma frequency \\
         $\nu_{\rm c}$& Cyclotron frequency \\
         $\nu_{\rm uh}$ & Upper hybrid frequency\\
        $\omega_{\rm p}$& Angular plasma frequency ($=2\pi\nu_{\rm p}$) \\
         $\omega_{\rm c}$& Angular cyclotron frequency ($=2\pi\nu_{\rm c}$)\\
         $\omega_{\rm uh}$ & Angular upper hybrid frequency ($=2\pi\nu_{\rm uh}$)\\
         $\theta$ & Angle between electromagnetic wave vector and magnetic field \\
         $\theta^\prime$ & Angle between Langmuir wave vector and magnetic field\\
         $h_{\rm p}$& Density scale height \\
         $k_{\rm B}$&Boltzmann's constant \\
         $c$&Speed of light \\
         $v_e$ & Electron thermal velocity \\
         $v_0$ & Hot plasma electron speed (cyclotron emission)\\
         $v_1$ & Hot plasma electron thermal speed (plasma emission)\\
         $\beta$ & velocity in units of $c$\\
         $\beta_{0.2}$ & $\beta/0.2$\\
         $\Gamma$ & Maser growth rate\\
         $L$ & Length scale of the magnetic trap\\
         $m_e$ & Electron mass \\
         $W_{\rm L}^{\rm tot}$ & Langmuir wave energy density\\
         $w$ & Fractional kinetic energy in turbulence\\
         $F$ & Flux-density\\
         $f$ & Fraction of stellar disc filled with flaring loops\\
         $s$ & Cyclotron harmonic number\\
         $T$ & Plasma temperature\\
         $T_1$ & Hot plasma component temperature\\
         $T_{\rm b}^{\rm L}$ & Langmuir wave's brightness temperature\\
         $T_{\rm b,h}$ & Brightness temperature of harmonic plasma emission\\
         $T_{\rm b,f}$ & Brightness temperature of fundamental plasma emission\\
         $T_{\rm b}^{\rm obs}$ & Observed brightness temperature\\
         $T_{\rm b}^{\rm max}$ & Peak brightness temperature of cyclotron maser emission\\
         $T_{\rm b}^{\rm avg}$ & Average brightness temperature of cyclotron maser emission\\
         $\tau_{\rm L}$ & Timescale for angular scattering of Langmuir waves\\
         $W_k$ & Spectral energy density of Langmuir waves\\
         $\Omega_{\rm L}$ & Beam solid angle of Langmuir waves\\
         $o,x$ & Ordinary and extraordinary magnetoionic modes\\
         $\tau_{o,x}$ & Cyclotron optical depth\\
         $L_{\rm B}$ & Magnetic scale length\\
         $\mu_{o,x}$ & Magnetoionic refractive index\\
         $T_{o,x}$ & Transverse magnetoionic axial ratio\\
         $K_{o,x}$ & Longitudinal magnetoionic axial ratio\\
         $\lambda_{\rm D}$& Plasma Debye length $=v_e/\omega_{\rm p}$\\
         $r_0$ &Classical electron radius\\
         $L_X$ & Soft X-ray luminosity ($\sim 0.1$ to $\sim 10$\,keV)\\
         $\alpha_{\rm f,h}$ & Fundamental and second harmonic emissivity for plasma emission\\ 
    \end{tabular}
    \label{tab:glossary}
\end{table*}

\section{Gyro-harmonic optical depth}
Although cyclotron maser emission is emitted by mildly relativistic electrons, the thermal background electrons can resonantly absorb the emitted waves at harmonics of the cyclotron frequency. The emitted radiation must necessarily pass through layers with decreasing field strength where absorption at the higher harmonics of the cyclotron frequency (gyro-harmonic absorption hereafter) can be catastrophic for radiation escape. 
The optical depth for the $o$- and $x$-mode are given by  \citep{stepanov-1995}
\begin{equation}
\tau_{o,x} = \pi\left(\frac{\omega_{\rm p}}{\omega_{\rm c}}\right)^2\frac{\omega L_{\rm B}}{c}\frac{s^{2s-2}}{s!}\left(\frac{2k_{\rm B}T\sin\theta}{m_ec^2}\right)^{2s-2}\,C_{o,x},
\end{equation}
where $s\geq2$ is the harmonic at which the optical depth is being computed, $L_B=B|\Delta B|^{-1}$ is the magnetic scale length, $m_e$ is the electron mass, $T$ is the plasma temperature, and $C_{o,x}$ is a mode dependent factor:

\begin{equation}
C_{o,x} = \mu_{o,x}^{2s-3}\frac{\left(1+T_{o,x}\cos\theta+K_{o,x}\sin\theta\right)^2}{1+T_{o,x}^2}.
\end{equation}
Here $\mu_{o,x}$ is the refractive index, $T_{o,x}$ and $K_{o,x}$ are the transverse and longitudinal axial ratios. They are given by
\begin{eqnarray}
T_o=-T_x^{-1} &=&\frac{Y(1-X)\cos\theta}{0.5Y^2\sin^2\theta-\Delta} \nonumber\\
\Delta &=& 0.25Y^4\sin^4\theta + (1-X)^2Y^2\cos^2\theta \nonumber \\
K_{o,x} &=& \frac{XY\sin\theta}{1-X}\frac{T_{o,x}}{T_{o,x}-Y\cos\theta}\nonumber \\
\mu^2_{o,x} &=& 1-\frac{XT_{o,x}}{T_{o,x}-Y\cos\theta}, 
\end{eqnarray}
where $X=\nu^2_{\rm p}/\nu^2$, $Y=\nu_{\rm c}/\nu$.
We will use the above expression for $C_{o,x}$ in our computations, but for a quick estimate one can use $C_{o,x}\approx \sqrt{\pi/2}\left[0.5(1-\sigma|\cos\theta|)\right]^2$ where $\sigma=1$ for the $o$-mode and $\sigma=-1$ for the $x$-mode.
\begin{figure*}
\label{fig:ghabs}
\centering
\begin{tabular}{ll}
\includegraphics[width=0.47\linewidth]{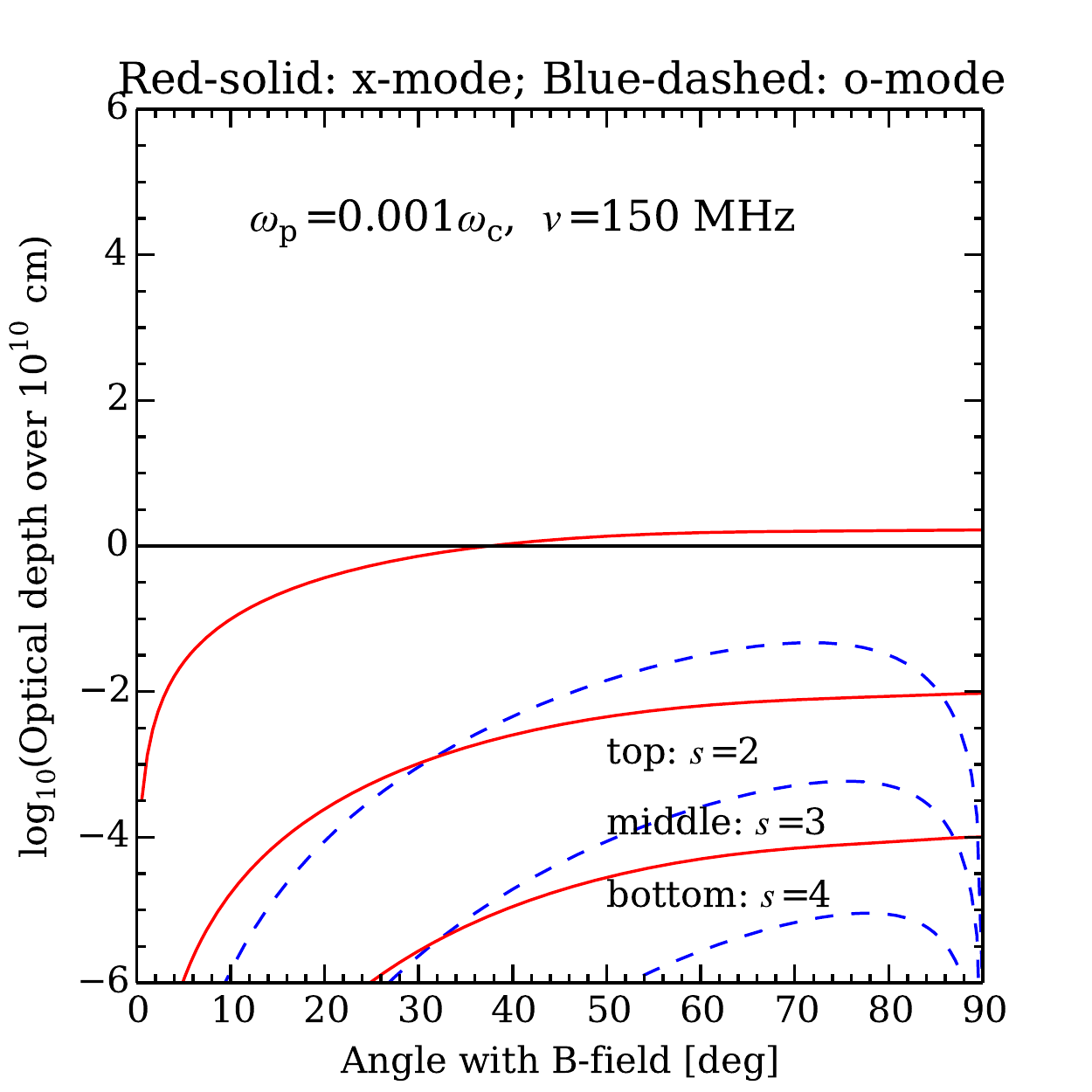} &
\includegraphics[width=0.47\linewidth]{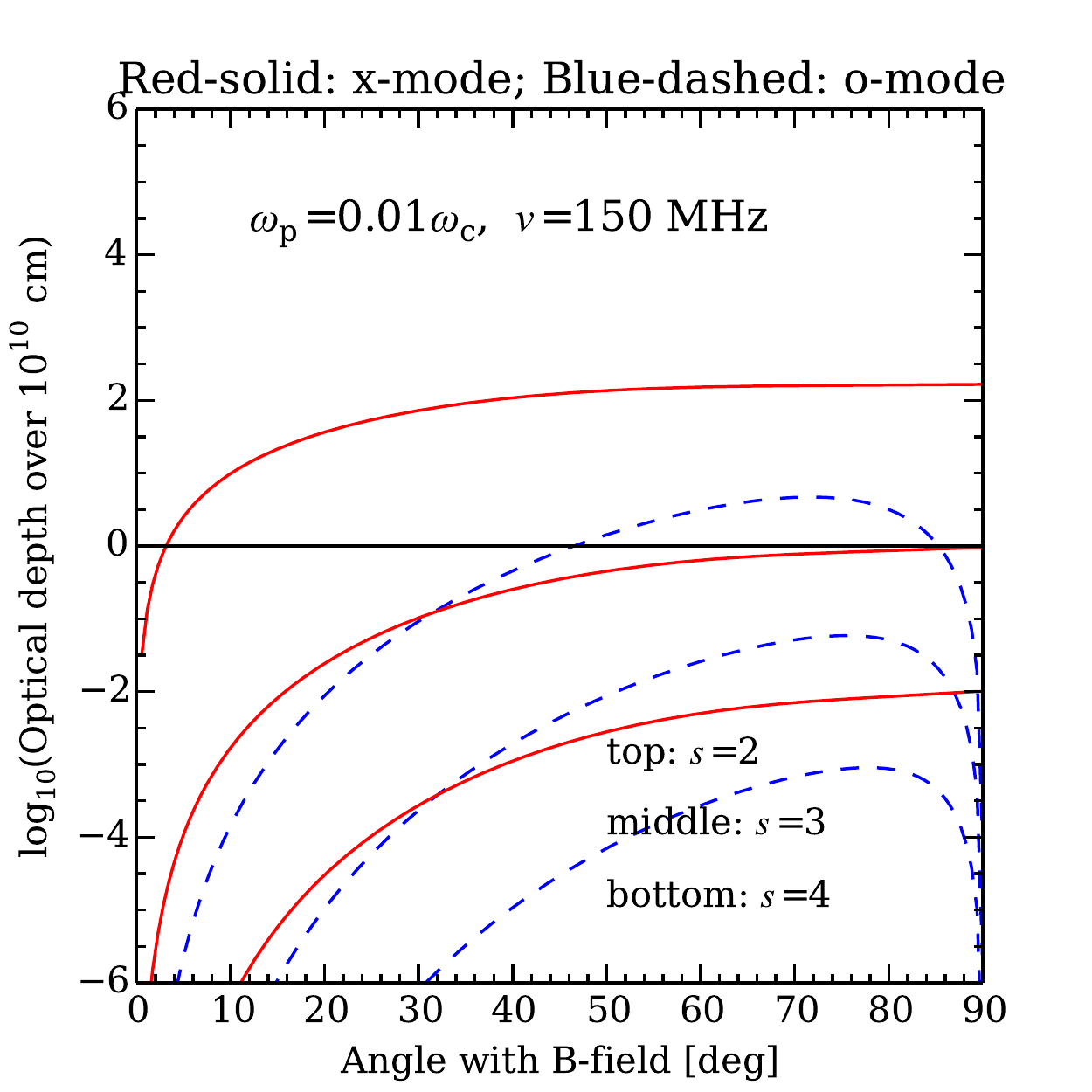} \\
\includegraphics[width=0.47\linewidth]{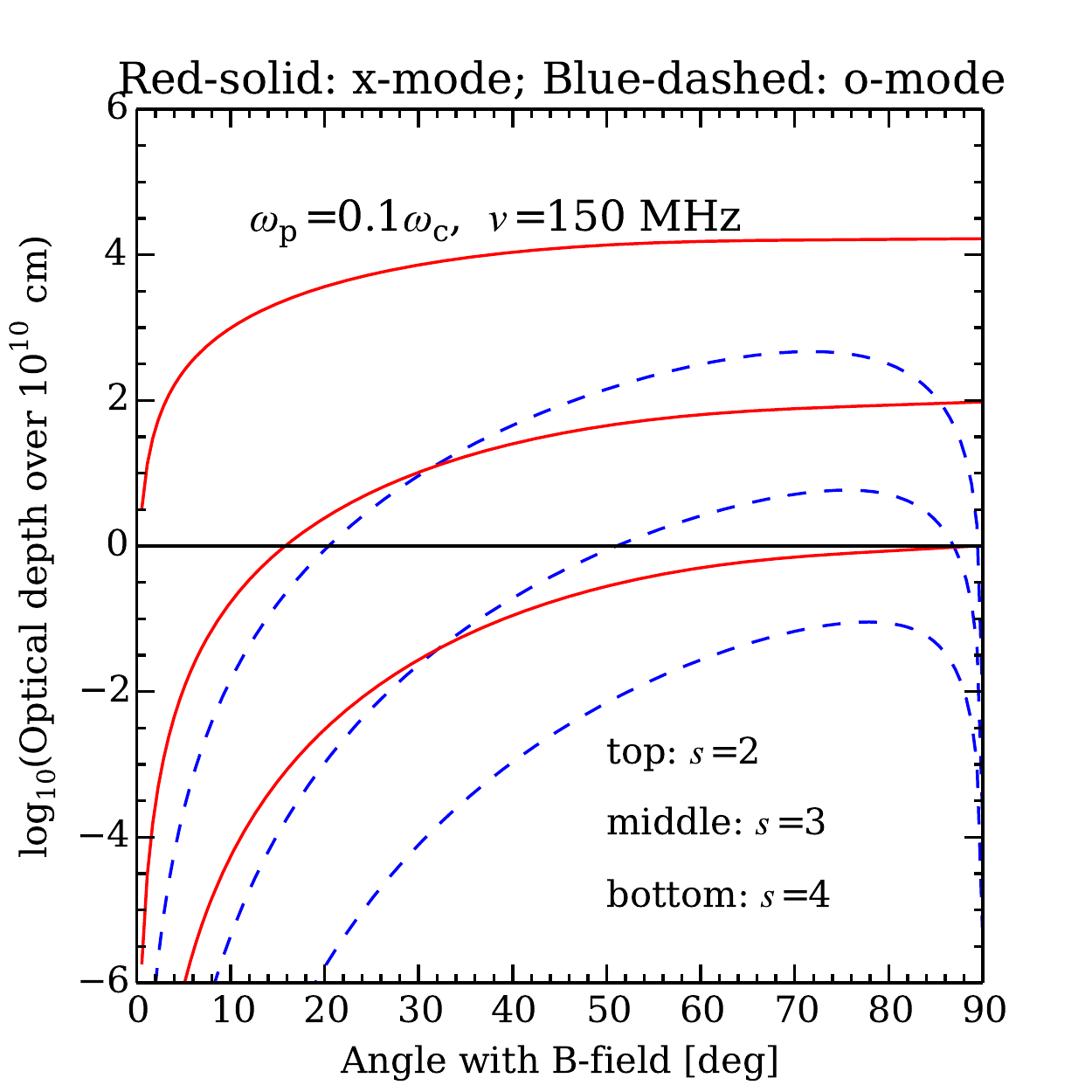} &
\includegraphics[width=0.47\linewidth]{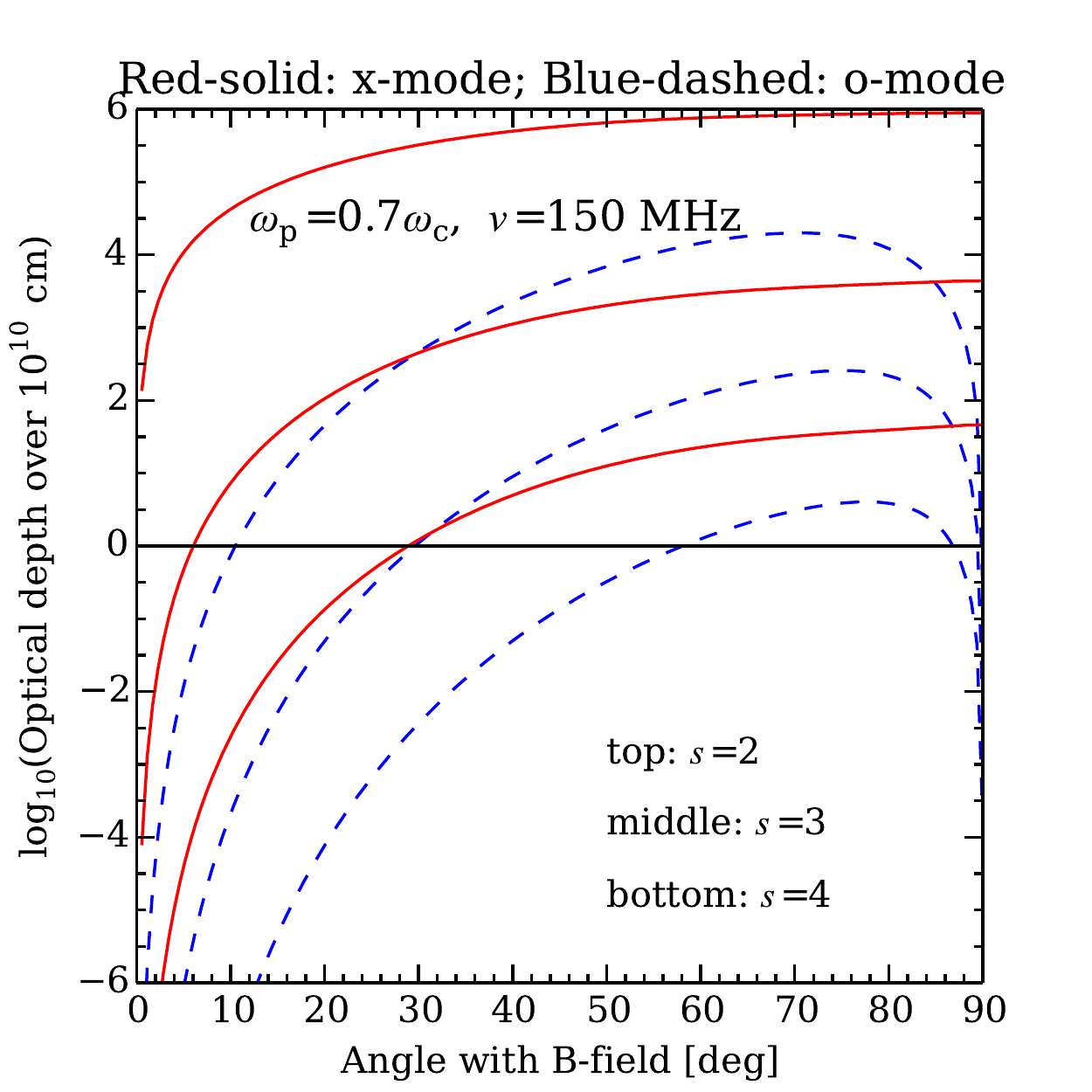} \\
\end{tabular}
\caption{Gyro-harmonic optical depth at different harmonics of the cyclotron frequency for the $x$- and $o$-mode radiation as a function of the angle between the wave-vector and the ambient magnetic field. Each panel assumes a different plasma to cyclotron frequency ratio. All panels assume radiation at $\nu=150\,{\rm MHz}$ and a scale length of $L=10^{10}\,{\rm cm}$. The optical depth scales as $\tau\propto L\nu(\nu_{\rm p}/\nu_{\rm c})^2$. }
\end{figure*}
Figure \ref{fig:ghabs} shows the gyro-harmonic optical depth at $150\,{\rm MHz}$ as a function of the angle between the wave-vector and the magnetic field, $\theta$, and plasma to cyclotron frequency ratio, $\nu_{\rm p}/\nu_{\rm c}$ . The optical depths have been computed for propagation over $L=10^{10}\,{\rm cm}$ which is comparable to the density scale heights of stellar coronae. Because the radiation is emitted at $\theta\sim \pi/2$, emission at the fundamental is near completely absorbed in the $s=2$ layer unless $\nu_{\rm p}/\nu_{\rm c}\lesssim 10^{-3}$ which for emission at $150\,{\rm MHz}$ corresponds to an ambient thermal plasma density of just $\sim 300\,{\rm cm}^{-3}$. Emission at the second harmonic will suffer catastrophic absorption at the third harmonic layer unless $\nu_{\rm p}/\nu_{\rm c}\lesssim 10^{-2}$ which for emission at 150\,MHz corresponds to a thermal plasma density of $\approx 7000\,{\rm cm}^{-3}$. For comparison, typical base pressure in the solar corona range between $10^8$ and $10^9\,{\rm cm}^{-3}$. 
The energy density in a cyclotron maser at metre-wavelengths can be extremely high ($T_{\rm b}\gtrsim 10^{14}\,{\rm K}$). This allows escape of radiation at $\gtrsim 10^{12}\,{\rm K}$ brightness temperature despite moderate levels of gyro-harmonic optical depth. As such a conservative limit for significant radiation to escape would be $\nu{\rm p}/\nu{\rm c}<0.01$. Incidentally, $\nu_{\rm p}/\nu_{\rm c}<0.01$ also corresponds to the absolute limit of $(\nu_{\rm p}/\nu_{\rm c})^2<0.5(1-\gamma^{-1})$ at which a shell-type maser with $\beta=0.2$ electrons is theoretically feasible \citep[equation 1 of][]{ergun2000}.
\section{Low-frequency approximation for the brightness of plasma emission }
Non-thermal growth of Langmuir waves is feasible with kinetic instabilities such as a bump-on-tail or loss-cone instability that typically yield a flat-spectrum  \citep[$W_{k^\prime}$ independent of $k^\prime$][]{kaplan1973} over a range of wavenumbers: $k^\prime\in[k^\prime_{\rm min},\,k^\prime_{\rm max}]$. In case of a loss-cone instability that likely operates in flaring loops, $k^\prime_{\rm min} = \omega_{\rm p}/v_1$ and $k^\prime_{\rm max} = \omega_{\rm p}/(3v_e)$ \citep{zaitsev1983}. The former is obtained from the condition for resonance that requires the phase velocity of Langmuir waves to be equal to the velocity of hot electrons. The latter is a necessary to ensure that Landau damping by thermal electrons does not arrest wave growth. 

The bandwidth of emission can be derived from the dispersion relationship for Langmuir waves: $\omega_{\rm L}^2 = \omega_{\rm p}^2(1+3k^{\prime 2}\lambda_{\rm D}^2)$, where $\lambda_{\rm D} = v_e/\omega_{\rm p}$ is the Debye length. The resulting bandwidth is
\begin{eqnarray}
\label{eqn:plasma_bw}
    \frac{\Delta\nu}{\nu} &\approx& \left(1+3k^{\prime 2}_{\rm max}\lambda_{\rm D}^2\right)^{1/2} - \left(1+3k^{\prime 2}_{\rm min}\lambda_{\rm D}^2\right)^{1/2} \nonumber \\ 
    &\approx& \frac{3}{2}\lambda_{\rm D}^2\left(k^{\prime 2}_{\rm max} - k^{\prime 2}_{\rm min}\right)
\end{eqnarray}

The emissivity for the fundamental and harmonic are given by 

\begin{equation}
\label{eqn:alpha_f}
    \alpha_{\rm f} = \frac{\pi}{36}\frac{m_ev_1^2}{k_{\rm B}}\frac{w\omega_{\rm p}^2}{\sqrt{3}k^\prime v_ec}
\end{equation}
and
\begin{equation}
\label{eqn:alpha_h}
    \alpha_{\rm h} = \frac{(2\pi)^5}{15\sqrt{3}}\frac{c^3}{\omega_{\rm p}^2v_1}\frac{w^2n_eT}{\xi^2}
\end{equation}
where the fractional power in turbulence, $w=W_{\rm L}^{\rm tot}/(nk_{\rm B}T)$ ($W_{\rm tot}^{\rm L}$ is the energy density of Langmuir waves), and the spectral bandwidth $\xi = (4\pi/3)\lambda_{\rm D}^3(k^3_{\rm max}-k^3_{\rm min})$. Note that the units of $\alpha_{\rm f}$ and $\alpha_{\rm h}$ above are Kelvins of brightness temperature per unit length. To get the brightness temperature we must multiply this emissivity with the length-scale over which the ambient plasma frequency does not vary by more than the instantaneous bandwidth of emission. Hence $T_{\rm b,f} = 2\alpha_{\rm f,h}h_{\rm p}\Delta\nu/\nu$ which yield equations \ref{eqn:plasma_tbfmax} and \ref{eqn:plasma_tbhmax}.

Finally, the brightness temperature of an isotropic distribution of Langmuir waves is given by $T^{\rm L}_{\rm b}k_{\rm B} = (2\pi)^3W_{k^\prime}/k^{\prime 2}$, where $\int \, {\rm d}k^\prime W_{k^\prime}$ is the volume energy density of Langmuir waves per unit solid angle. For an isotropic flat spectrum of Langmuir-wave turbulence $W_{k^\prime}$ is independent of $k^\prime$ and we have $4\pi W_{k^\prime}(k^\prime_{\rm max}-k^\prime_{\rm min}) = W_{\rm L}^{\rm tot} = w/(nk_{\rm B}T)$. To get a conservative order of magnitude upper bound on the  brightness of Langmuir waves, we set $w=10^{-5}$, $k^\prime \approx k^\prime_{\rm min} = \omega_{\rm p}/v_1$, $k^\prime_{\rm max}-k^\prime_{\rm min}\approx k^\prime_{\rm max} = \omega_{\rm p}/(3v_e)$ to get
\begin{equation}
    T^{\rm L}_{\rm b} \lesssim  \frac{3}{4\pi}\frac{nTwv_ev_1^2}{\nu_{\rm p}^3} 
\end{equation}
which gives equation \ref{eqn:tb_max_lang}.
\section{Polarization of second harmonic plasma emission}
Theoretical investigation into the polarisation of transverse emission at the second harmonic due to coalescence of two Langmuir waves can be found in \citet{mds-2pol,zlotnik1981,melrose-2harmpol}. Here we make use of the formulae given by \citet[][their equations 1 through 4]{melrose-2harmpol}. The degree of isotropy of the Langmuir wave spectrum enters the equations through the factors $A$, $B$, and $C$ in their equations 2a through to 2c. For the purely isotropic case, we have $\left<\cos^0\theta^\prime\right>=2$, $\left<\cos^2\theta^\prime\right>=2/3$ and $\left<\cos^4\theta^\prime\right>=2/5$, which gives $A\approx 0.133$, $B=0$ and $C=0$.  

Higher polarisation fractions are obtained for higher values of $\nu_{\rm c}/\nu_{\rm p}$. However some care is needed in interpreting emission and propagation issues where the ratio is of order unity. This is due to two factors: (a) Gyro-harmonic absorption may preclude detectable radiation from escaping the source for $\nu_{\rm c}\sim \nu_{\rm p}$. Additionally,  the Langmuir waves are actually excited not just above the plasma frequency but rather the upper hybrid frequency given by $\omega_{\rm uh} \approx \sqrt{\omega^2_{\rm p} + \omega^2_{\rm c}\sin^2\theta}$. (b) A large magnetic field may prevent Langmuir waves from achieving isoptropy thereby modifying the polarisation properties of the emission.

From Fig. \ref{fig:ghabs}, it is clear that for $\nu_{\rm c}/\nu_{\rm p}$ much larger than $0.1$, even the $s=4$ layer can have prohibitive levels of absorption. We therefore enforce the condition $2\nu_{\rm uh} > 4\nu_{\rm c}$ (factor of 2 is for second harmonic emission) to avoid catastrophic gyro-harmonic absorption, which gives $\nu_{\rm c}/\nu_{\rm p}<(4-\sin^2\theta)^{-1/2}$. The R.H.S varies between 0.5 and 0.577 as $\theta$ increases from 0 to $\pi/2$. We note that due to the low $x$-mode gyro-harmonic absorption at $\theta\approx 0$, somewhat higher values of $\nu_{\rm c}/\nu_{\rm p}$ may lead still lead to escaping radiation but this would require a contrived geometry. While this is certainly feasible in some cases where the star is viewed pole-on, but it cannot be true of the vast majority of observed events. 

The non-isotropic spectrum of Langmuir waves in the presence of a magnetic field has been discussed by \citet{mds-2pol}. The growing wave-spectrum is confined to forward and backward cones of opening angle $\theta^\prime \lesssim \sin^{-1}(\sqrt{3}k^\prime v_e/\omega_{\rm c})$. Letting the Langmuir wavenumber $k^\prime = k_{\rm max} = \omega_{\rm p}/(3v_e)$, we get $\theta^\prime \lesssim \sin^{-1}\left[\nu_{\rm p}/(\sqrt{3}\nu_{\rm c})\right]$. Within the gyro-harmonic constraint of $\nu_{\rm c}/\nu_{\rm p}\lesssim 0.577$, the argument of $\sin^{-1}$  is larger than unity, implying that the field does not restrict waves from growing isotropically. For $\nu_{\rm c}/\nu_{\rm p}\gtrsim 0.6$ however, wave growth is restricted, but escape from gyro-harmonic absorption is only possible at low $\theta$. We have denoted by using dotted lines in Fig. \ref{fig:2harmpol}.

\bibliographystyle{mnras}

\bsp	
\label{lastpage}
\end{document}